\documentclass[aps,twocolumn,graphics,floatfix,tightenlines]{revtex4-1}
\usepackage{graphics}
\usepackage{epstopdf}
\usepackage{epsfig}
\usepackage{graphicx}
\usepackage{epsf,epic}
\usepackage{color}
\usepackage[flushleft]{threeparttable}
\usepackage{subfig}
\usepackage{amsmath}
\usepackage{booktabs}
\usepackage{multirow}
\usepackage{physics}
\usepackage{hyperref}
\usepackage{amsfonts}
\usepackage{wrapfig}
\usepackage{breqn}
\usepackage{pstricks}
\usepackage[utf8x]{inputenc}
\usepackage{multirow}
\usepackage{fancyref}
\usepackage{pst-node}
\usepackage{float}
\usepackage{bm}
\usepackage{dcolumn}

\makeatletter
\usepackage{etoolbox} 
\appto{\appendix}{%
	\@ifstar{\def\theequation@prefix{A.}}%
	{}%
}
\preto\maketitle{%
  \begingroup\lccode`~=`,
  \lowercase{\endgroup
  \let\saved@breqn@active@comma~
  \let~}\active@comma 
}
\appto\maketitle{%
  \begingroup\lccode`~=`,
  \lowercase{\endgroup
  \let~}\saved@breqn@active@comma 
}
\makeatother
\begin{document}
\title{Quaternary MgSiN$_2$-GaN alloy semiconductors for deep UV applications}
\author{Ozan Dernek}\email{ozan.dernek@case.edu}
\author{Walter R. L. Lambrecht}\email{walter.lambrecht@case.edu}
\affiliation{Department of Physics, Case Western Reserve University, 10900 Euclid Avenue,
             Cleveland, OH-44106-7079}

\begin{abstract}
  Ultra-wide direct band gap semiconductors hold great promise for deep ultraviolet
  opto-electronic applications. Here we evaluate the potential of MgSiN$_2$--GaN alloys
  for thus purpose. Although MgSiN$_2$ itself has an indirect gap $\sim$0.4 eV below its
  direct gap of $\sim$6.5 eV, its different sign lattice mismatch from GaN in two
  different basal plane directions could avoid the tensile strain which limits
  Al$_x$Ga$_{1-x}$N on GaN for high $x$. Two octet-rule preserving structures (with
  space groups $Pmn2_1$ and $P1n1$) of a 50 \% alloy of MgSiN$_2$ and GaN are investigated
  and are both found to have gaps larger than 4.75 eV using quasiparticle self-consistent
  (QS) $GW$ calculations. Both are nearly direct gap in the sense that the indirect gap is
  less than 0.1 eV lower than the direct gap. Their mixing energies are positive yet
  small, with values of 8 (31) meV/atom for $Pmn2_1$ ($P1n1$) indicating only a small
  driving force toward phase separation. 
\end{abstract}
\maketitle

\section{Introduction}
  Compact light sources in the deep UV, in particular the UV-B and UV-C with wavelengths
  shorter than 315 nm and 280 nm respectively would have a large impact in science and
  technology \cite{Kneissl2019}. Light emitting diodes (LED) in this range could fulfill
  this demand but require new semiconductor materials with efficient doping and band gaps
  $E_g>4$ eV. The current prevailing approach toward this goal is to develop
  Al$_x$Ga$_{1-x}$N alloys, which form a continuous alloy system with wurtzite structure
  and end point band gaps of 3.6 eV (GaN) and 6.3 eV (AlN).
  \cite{Crawford2017,Detchprohm2017,Shatalov2017, Hirayama2017,Zhao2017} However, high
  Al-content alloys suffer from a number of challenges: (1) epitaxial growth on GaN
  substrates results in tensile strain, which tends to develop cracks, (2) n-type doping
  with Si leads to distorted DX center with a deep level and hence inefficient doping,
  (3) p-type doping is even more problematic and (4) the inverted crystal field splitting
  of AlN leads to predominantly transverse magnetic (TM) light emission from the basal
  plane surface.\cite{Zhang2010,Reich2015} Here we propose alloys of MgSiN$_2$ and GaN as
  an alternative to overcome at least some of these problems.

  Heterovalent  ternary semiconductors of the type II-IV-N$_2$ using group II and group-IV
  element pairs as a replacement for the group III element in III-N nitrides, have gained
  significant interest in recent years.\cite{Lambrechtbook,ictmc21,Martinez17} These
  materials have a wurtzite derived structure, which ideally consists in a fully ordered
  orthorhombic superlattice of wurtzite with space group $\#33$ ($Pbn_1$ in the setting
  with $a_o\approx2a_w, b_o\approx\sqrt{3}a_w,c_o=c_w$ where $a_w,c_w$ are the wurtzite
  and $a_o, b_o,c_o$ the orthorhombic lattice constants, or $Pna2_1$ if $a_o$ and $b_o$
  are reversed). They add significant flexibility to the chemical materials design space
  because one has a choice of II=Be, Mg, Zn, Cd and IV= Si, Ge, Sn. Furthermore, a certain
  degree of polytypic disorder with space group $\#26$ ($Pmc2_1$) which also satisfies the
  octet rule \cite{Quayle15}, or the presence of exchange defects II$_{\rm IV}$ and
  IV$_{\rm II}$ can be tolerated and allows for some disorder tuning of the gap
  \cite{Skachkov16x,Lany17,Cordell22,Veal15}.

  The design flexibility is further enhanced by considering mixed II-IV-N$_2$ with III-N
  alloy systems. Recently, the ZnGeN$_2$-GaN alloy system was investigated both
  computationally\cite{Dinushi18} and experimentally in thin film
  growth\cite{Dinushi20,Suehiro2017}. In particular, it was found that at 50\% two
  octet-rule preserving structures can be constructed theoretically with space groups $Pmn2_1$ and $P1n1$, of which the
  former was found to have the lowest total energy. In the $Pmn2_1$ structure, each N is
  surrounded by exactly two Ga, one Zn and one Ge and the structure can be viewed as half
  a layer of ZnGeN$_2$ and half a layer of GaN in the $b$ direction or a $[010]_{1/2}$
  lattice. Experimentally alloys at the 50 \% composition were achieved\cite{Dinushi20} but their detailed structure
  and degree of ordering is yet to be confirmed. We point out that a line compound at 50 \%
  has been found to exist\cite{Omata2011} in the analogous 50 \% mixed compound of ZnO and LiGaO$_2$,
  which is a mixture of  the parent II-VI compound and its ternary I-III-VI$_2$.

  Here we use this flexibility to search for a semiconductor which is closely lattice
  matched to GaN and has a gap exceeding 4 eV. From, the band gap {\sl vs.} equivalent
  wurtzite lattice constant diagram in Ref. \onlinecite{ictmc21}, one can see that
  MgSiN$_2$ is significantly closer lattice matched to GaN than AlN and has a predicted
  direct gap of $\sim$6.3-6.5 eV \cite{Atchara16,Quirk14}. Unfortunately, its indirect gap is
  about 0.4 eV lower, so it is actually an indirect gap semiconductor which is undesirable
  for light emission devices. Upon closer inspection of the lattice match, one should note
  that the equivalent wurtzite lattice constant in \cite{ictmc21} was taken as $a_o/2$.
  However, inspecting the individual orthorhombic lattice constants, there is only a 1.1\%
  mismatch of MgSiN$_2$ with respect to 2a$_{\rm GaN}$ in the $a$-direction, but a
  $-4.7$\% mismatch in the $b$ direction and a $-4$ \% mismatch in the $c$-direction.
  Nonetheless, these can be viewed as acceptable lattice mismatch compared with AlN which
  has a $-2.4,-3.9$ \% mismatch in $a$ and $c$ directions compared to GaN. In particular,
  a compressive strain in the basal plane would be less disadvantageous than a tensile
  strain and at least in one basal plane direction, the strain of MgSiN$_2$ on GaN would
  be compressive. 

  The indirect gap in MgSiN$_2$ results from the valence band maximum (VBM) occurring at
  $U=(\pi/a,0,\pi/c)$ in the $Pbn2_1$ setting of the space group, while the conduction band
  minimum (CBM) occurs at $\Gamma=(0,0,0)$. Considering the gap of GaN at $U$ when
  calculated in an equivalent $Pbn2_1$ supercell, which lies 1.1 eV below the actual VBM
  at $\Gamma$, and interpolating the direct $\Gamma-\Gamma$ and indirect $U-\Gamma$ gaps
  linearly between MgSiN$_2$ and GaN, using a direct(indirect) gap of 3.4(4.5) eV in GaN
  and (6.5,6.1) in MgSiN$_2$ \cite{Atchara16} one might expect a direct gap up to about
  75 \% and a direct gap of 5.8 eV at that point. Of course, alloy band gaps typically
  show band gap bowing and the above also ignores strain effects. Nonetheless, this seemed
  sufficiently promising to investigate the band gap in a quaternary MgSiN$_2$-GaN system
  further.

\section{Computational Method}
  The stability of the 50 \% alloy is investigated by calculating first the cohesive
  energy {\sl vs.} the isolated atoms, second, the enthalpy  of formation with respect to
  the elements in their standard state, and finally, the mixing energy {\sl vs.} the two
  end compounds GaN and MgSiN$_2$. The total energies required for each system are
  calculated using density functional theory (DFT) within the Generalized Gradient
  Approximation (GGA) in the Perdew-Burke-Ernzerhof (PBE)\cite{PBE} parametrization. The
  structural parameters, internal position and lattice constants were fully relaxed using
  the {\sc Quantum Espresso} code \cite{QE-2009} within the Projector Augmented Wave (PAW)
  method. Subsequently they were also calculated using the all-electron {\sc Questaal}
  code which implements the full-potential linearized muffin-tin orbital (FP-LMTO) method.
  In evaluating the total energy differences between different systems, we took care of
  using exactly the same muffin-tin radii for the atoms in different systems and use
  exactly equivalent {\bf k}-point meshes and real space meshes to maximize systematic
  error cancellation. All, three systems, GaN, MgSiN$_2$ and MgSiGa$_2$N$_4$ were
  calculated in the same 16 atom cell to achieve equal convergence. The cohesive 
  energy calculations use the single atom energies calculated in large
  vacuum cells (with spin polarization for Ga, Si and N) and the most stable form of each
  species. The cohesive energies of the elements were obtained using both code suites and are compared to
  experimental values in Table~\ref{table:cohen}.

    \begin{table}
      \caption{Cohesive energies of each species calculated by {\sc Questaal} and
               {\sc QE}. Species Ga, Mg, and Si are calculated in their bulk phases, and
               N$_{\mathrm2}$ molecule calculated in a large vacuum cell.
               Spin-polarization is included for single atom calculations in {\sc QE}.
               \label{table:cohen}}
    \begin{ruledtabular}
      \begin{tabular}{lcccc}
        Species & Space Group    & \multicolumn{3}{c}{Coh. En. (eV/atom)} \\
                &                & {\sc QUESTAAL} & {\sc QE} & Expt.      \\\hline
        Ga      & $Cmce$           & -2.69          & -2.71    & -2.81      \\
        Mg      & $P6_3/mmc$     & -1.50          & -1.50    & -1.51      \\
        Si      & $Fd\bar{3}m$   & -4.56          & -4.62    & -4.63      \\
        N       & N$_{\mathrm2}$ & -5.04          & -5.18    & -4.92      \\
      \end{tabular}
     \end{ruledtabular}
     \end{table}
  
  The band structures were calculated using the quasiparticle self-consistent (QS) $GW$
  approach \cite{MvSQSGWprl,Kotani07} implemented in the {\sc Questaal}
  code\cite{questaalpaper}. Here $GW$ stands for the one-electron Green's function and $W$
  for the screened Coulomb interaction in Hedin's approach \cite{Hedin65}, which define
  the self-energy $\Sigma=iGW$. The difference between this non-local and energy dependent
  self-energy operator $\Sigma(\omega)$ and the exchange-correlation potential $v_{xc}$
  is, as usual, calculated within first-order perturbation theory starting from the DFT
  Kohn-Sham (KS) eigenvalues $\epsilon_i$ and eigenstates $\psi_i$. However, in the QS
  approach, a new energy independent but non-local potential, expressed by its matrix
  elements in the basis of KS eigenstates, $\tilde \Sigma_{ij}=\frac{1}{2}\mathrm{Re}[\Sigma_{ij}(\epsilon_i)+\Sigma_{ij}(\epsilon_j)]$ is evaluated and iterated to convergence.
  Here $\mathrm{Re}[\dots]$  means taking the hermitian part. The eigenvalues thereby
  become independent of the DFT starting point exchange-correlation choice. They are found
  to be typically somewhat overestimating band gaps because within standard $GW$ the
  screening of the Coulomb potential is calculated in the random phase approximation (RPA)
  which underestimates screening. This can be overcome by adding ladder
  diagrams\cite{Cunningham18,Cunningham21,Radha21} but, as a computationally less demanding  alternative, it is found
  that using 80\% of the $\tilde\Sigma$ and 20\% DFT exchange correlation potential
  usually gives very accurate results for a wide variety of
  systems\cite{Bhandari18,Deguchi2016}. We follow this approach here.

  The {\sc Questaal} implementation of the $GW$ approach uses a mixed interstitial plane
  wave, product basis set to expand two-point quantities, such as $W=(1-vP)^{-1}v$, with
  $v$ the bare Coulomb interaction and $P=-iGG$, the polarization propagator. This is a
  more efficient basis set than plane waves to represent the screening and, as a result,
  high-energy states of the KS equation are less crucial to the convergence of the method.
  The use of an atom-centered basis set allows one to express the self-energy
  $\tilde\Sigma$ in real space and hence to interpolate for {\bf k}-point different from
  the ones for which $\Sigma({\bf k},\omega)$ is calculated. This way, accurate band
  dispersions along the symmetry lines and effective masses can be obtained. We tested
  convergence of the various parameters entering the $GW$ calculations, finding a
  $3\times3\times3$ {\bf k}-mesh, $spdf-spd$ basis set including Ga-$3d$ local orbitals,
  and $E_{max}$ above which the self-energy is approximated by an average value to be
  well converged. Details of the QS$GW$ implementation can be found in Ref.\cite{Kotani07}.

\section{Results}
\begin{figure}
      \includegraphics[width=8cm]{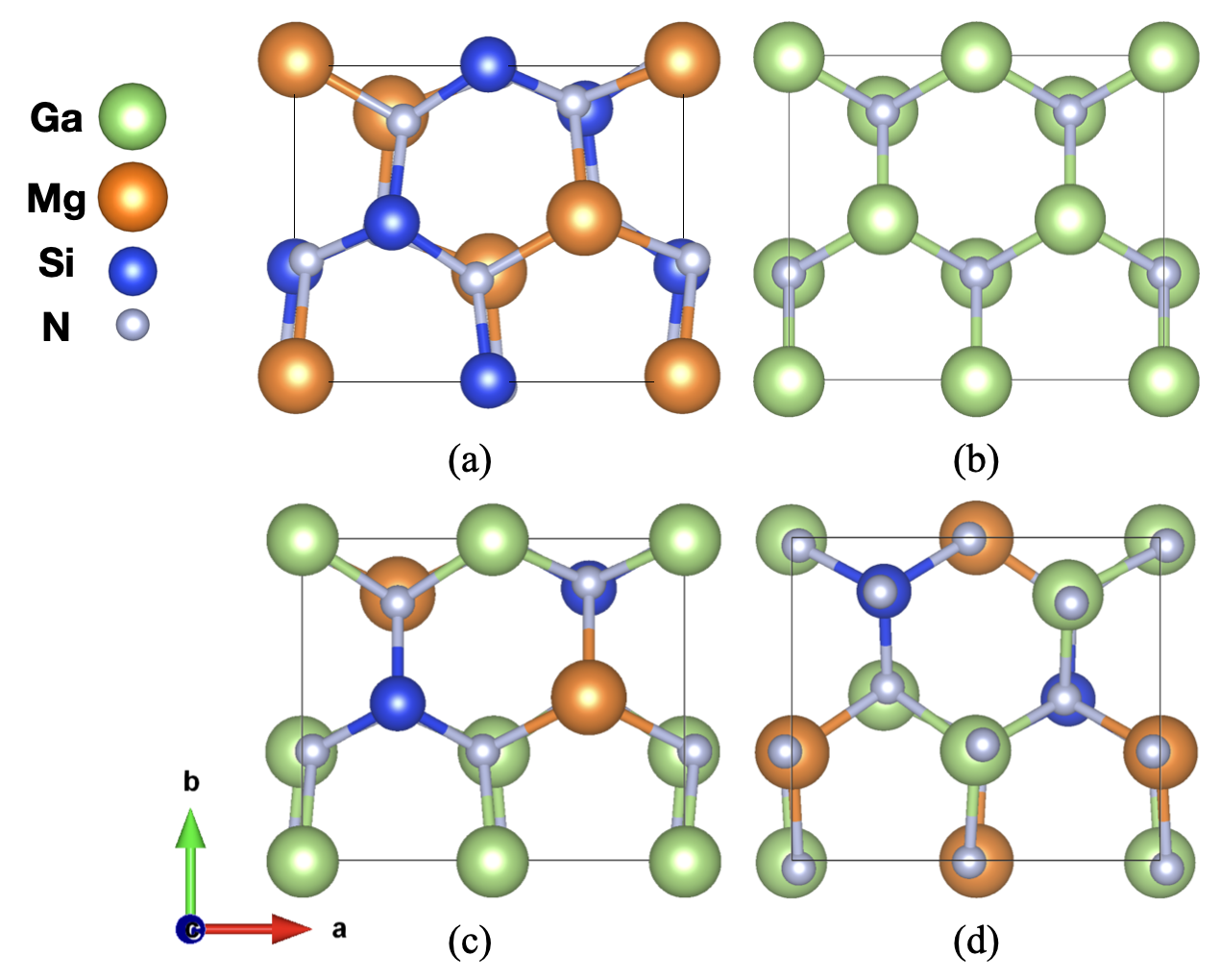}
      \caption{Octet-rule-preserving MgSiGa$_2$N$_4$ structures in 
         (c) $Pmn2_1$ and (d) $P1n1$ space groups and their relation to
                 the compounds (a) MgSiN$_2$ and (b) GaN both in the $Pbn2_1$ structure.
                 \label{fig:struc}}
    \end{figure}

  We here present results for both the $Pmn2_1$ and $P1n1$ structures, shown in Fig. \ref{fig:struc}, which were both
  found to have low energy in the ZnGeGa$_2$N$_4$ compound at 50\% mixing. We start with
  the structural parameters, given in Table \ref{table:lattice}. We compare the lattice
  constants with the Vegard's average. $Pmn2_1$ perfectly matches these values. When all
  the lattice constants and angles between them are relaxed, the $P1n1$ structure slightly
  shifts into a monoclinic phase, with the angles between the lattice vectors
  $\alpha=\gamma=90^{\circ}$ and $\beta=90.16^{\circ}$. This is consistent with the space group
  which belongs in the monoclinic crystal system.  
  However, the change in the volume
  and total energy and the total energy is negligible. Therefore we continue to
  investigate the material in its orthorhombic phase. The $P1n1$ lattice constant $a$
  falls  out of the range set by GaN and MgSiN$_2$, while $b$ and $c$ are very close to
  the GaN lattice constants. Yet the volume of the two structures are identical. The
  reduced coordinates for the atoms in the $Pmn2_1$ and for the $P1n1$ are provided in
  Appendix \ref{app-struc}.
  Compared to experimental values, we may notice that the volume is overestimated by 
  2.1 \% for MgSiN$_2$ and by 2.8 \% for GaN. We thus also expect an overestimate by
  a similar amount for the 50 \% compound. 
  
   \begin{table*}
     \caption{Average bond lengths and relaxed lattice constants in \AA\  for compounds
              and MgSiGa$_2$N$_4$ structures. Unit cell volumes are compared with the
              Vegard's average of MgSiN$_2$ and GaN.\label{table:lattice}}
      \begin{ruledtabular}
        \begin{tabular}{lccccccc}
       & Mg-N (\AA) &Ga-N (\AA) &Si-N (\AA) & $a$ (\AA) & $b$ (\AA) & $c$ (\AA) & Volume (\AA$^3$) \\
          \hline
MgSiN$_2$      & 2.11 && 1.76 & 6.504 & 5.310 & 5.031 & 173.75  \\
Expt.\footnote{R.J. Bruls, H. T. Hintzen,  R. Metselaar,  and  C. Loong  J. Phys.  Chem. Solids  61 1285.}      & 2.09 && 1.75 & 6.473 & 5.272 & 4.986 & 170.15 \\          
GaN            &      & 1.96  & & 6.433 & 5.571& 5.240 & 187.79  \\
Expt.\footnote{T. Detchprohm, K.Hiramatsu,  K. Itoh,  I. Akasaki  Jpn. J.  Appl. Phys.  (Japan),  vol. 25 (1986), p. L1454-6}      &      & 1.95  & & 6.378 & 5.523& 5.185 & 182.65  \\
$Pmn2_1$       & 2.06 & 1.95 &1.75 & 6.467 & 5.440 & 5.147 & 181.07  \\
$P1n1$         & 2.04 & 1.95 &1.75 & 6.324 & 5.550 & 5.159 & 181.07  \\
Vegard's       &      &&    & 6.469 & 5.441 & 5.136 & 180.75    \\
        \end{tabular}
  \end{ruledtabular}
    \end{table*}

    \begin{table}
      \caption{Band gaps of compounds and alloys in various approximations.
               \label{table:gap}}
      \begin{ruledtabular}
        \begin{tabular}{lcccc}
          Compound                   &          & GGA  & 0.8$\Sigma$   & 0.8$\Sigma$ + $\Delta(0)$\footnote{$\Delta(0)$ is an estimated zero-point motion correction from Ref. \cite{Atchara16}}\\\hline
                                                       
          MgSiN$_2$                  & Indirect & 4.03 & 6.15          & 5.84   \\
          MgSiN$_2$                  & Direct   & 4.36 & 6.52          & 6.28  \\
          GaN                        & Direct   & 1.72 & 3.08          &   \\
          MgSiGa$_2$N$_4$ ($Pmn2_1$) & Direct   & 2.98 & 4.83          &   \\
          MgSiGa$_2$N$_4$ ($Pmn2_1$) & Indirect & 2.92 & 4.76          &   \\
          MgSiGa$_2$N$_4$ ($P1n1$)   & Direct   & 2.73 & 4.60          &   \\
          MgSiGa$_2$N$_4$ ($P1n1$)   & Indirect & 2.70 & 4.57          &   \\
        \end{tabular}
      \end{ruledtabular}
    \end{table}
 
  Next, we discuss the total energy results which allow us to ascertain the stability of
  the proposed alloy. Table\ref{table:formation} lists the cohesive energies, enthalpies of formation
  and mixing energy all evaluated using both methods. Values obtained by the QE
  method are given in parentheses. The enthalpies of formation give the energy difference with
  respect to the elements in their standard state at room temperature and atmospheric
  pressure. Their negative value indicate stability. A more strict criterion is to
  evaluate the mixing energy of the alloy with respect to the two separate compounds.
  These values are positive but quite small. The positive value  indicates, as usual for semiconductor
  alloys, that there exists a miscibility gap and mixing can only occur above a
  certain temperature where entropy of mixing  makes the Gibbs free energy negative. But once the alloys are formed
  at the growth temperature, the atomically mixed frozen-in structure is expected to be kinetically stable at lower temperatures.  The miscibility gap temperature is directly related to the mixing energy at 50 \% composition and this being low
  indicates a relatively small driving force toward phase segregation and a low mixing temperature. 
  The energy of mixing reported here assumes perfect satisfaction of the octet rule and is thus
  a lower limit of the mixing enthalpy in a more realistic alloy structure, which may be expected to have a certain
  degree of disorder, in particular short-range disorder  involving local motifs around the N anion deviating from
  the perfect Ga$_2$MgSi by either excess Ga, excess Mg or excess Si. The study of these defects and their energy cost
  is a topic that needs to be addressed in future research.

        \begin{table}
      \caption{Cohesive energies, energy of formation, and mixing energies in GGA with
        units eV/atom. The values given in parenthesis are obtained by the QE method, the other ones
        by {\sc Questaal}.
               \label{table:formation}}
      \begin{ruledtabular}
        \begin{tabular}{lrrrr}
  & MgSiN$_2$\hspace{2mm} & GaN\hspace{5mm} & $Pmn2_1$\hspace{2mm} & $P1n1$\hspace{4mm} \\
        \hline
E$_{coh}$ &  5.25 (5.31)  &  4.39 (4.42)    &  4.82 (4.86)         &  4.80 (4.84)\\
E$_{for}$ & -1.19 (-1.19) & -0.52 (-0.47)   & -0.90 (-0.83)        & -0.88 (-0.81)\\
E$_{mix}$ &               &                 & 0.008 (0.007)        &  0.029 (0.031)
        \end{tabular}
      \end{ruledtabular}
    \end{table}

\begin{figure}
      \includegraphics[width=8cm]{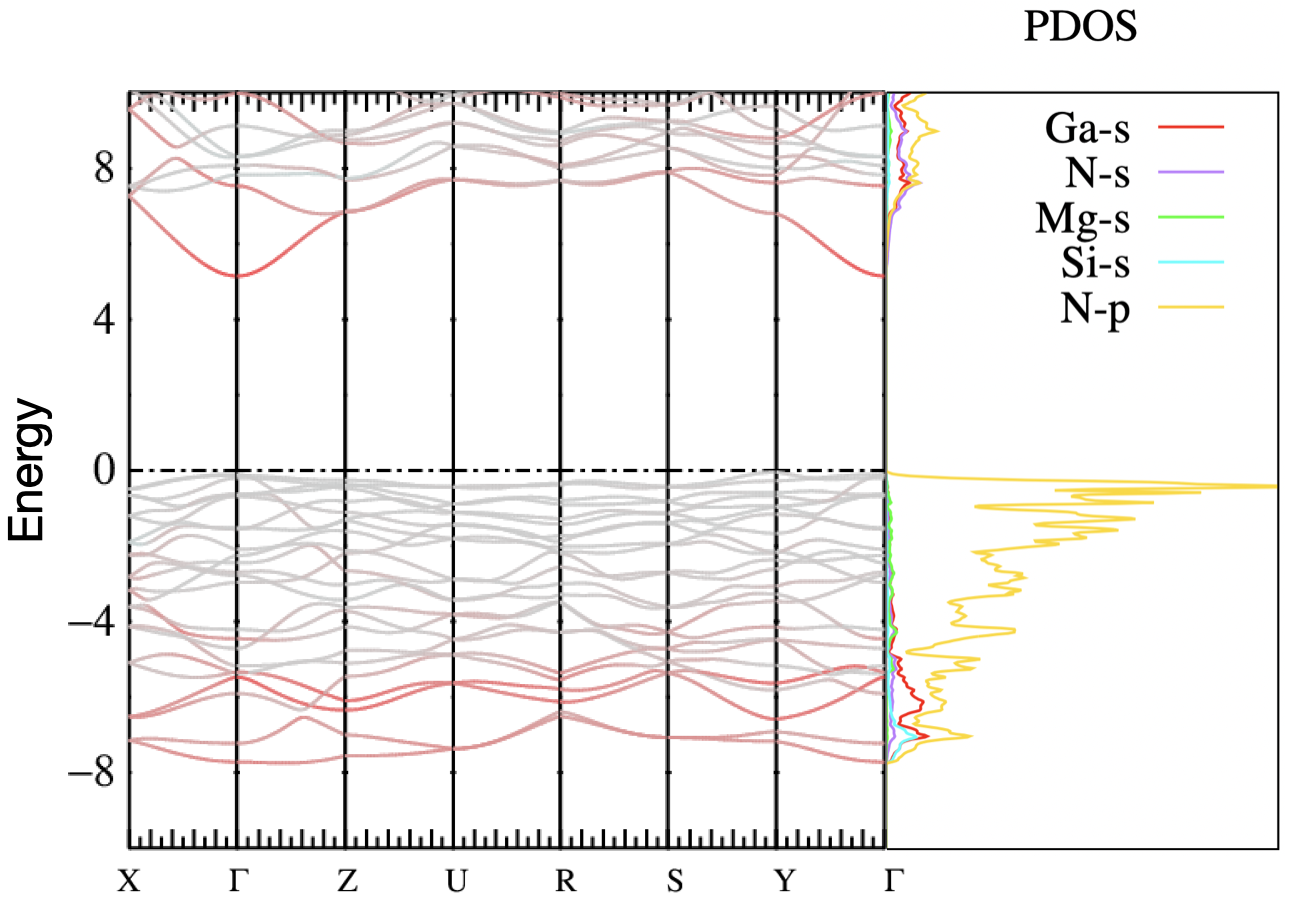}\\
      \includegraphics[width=8cm]{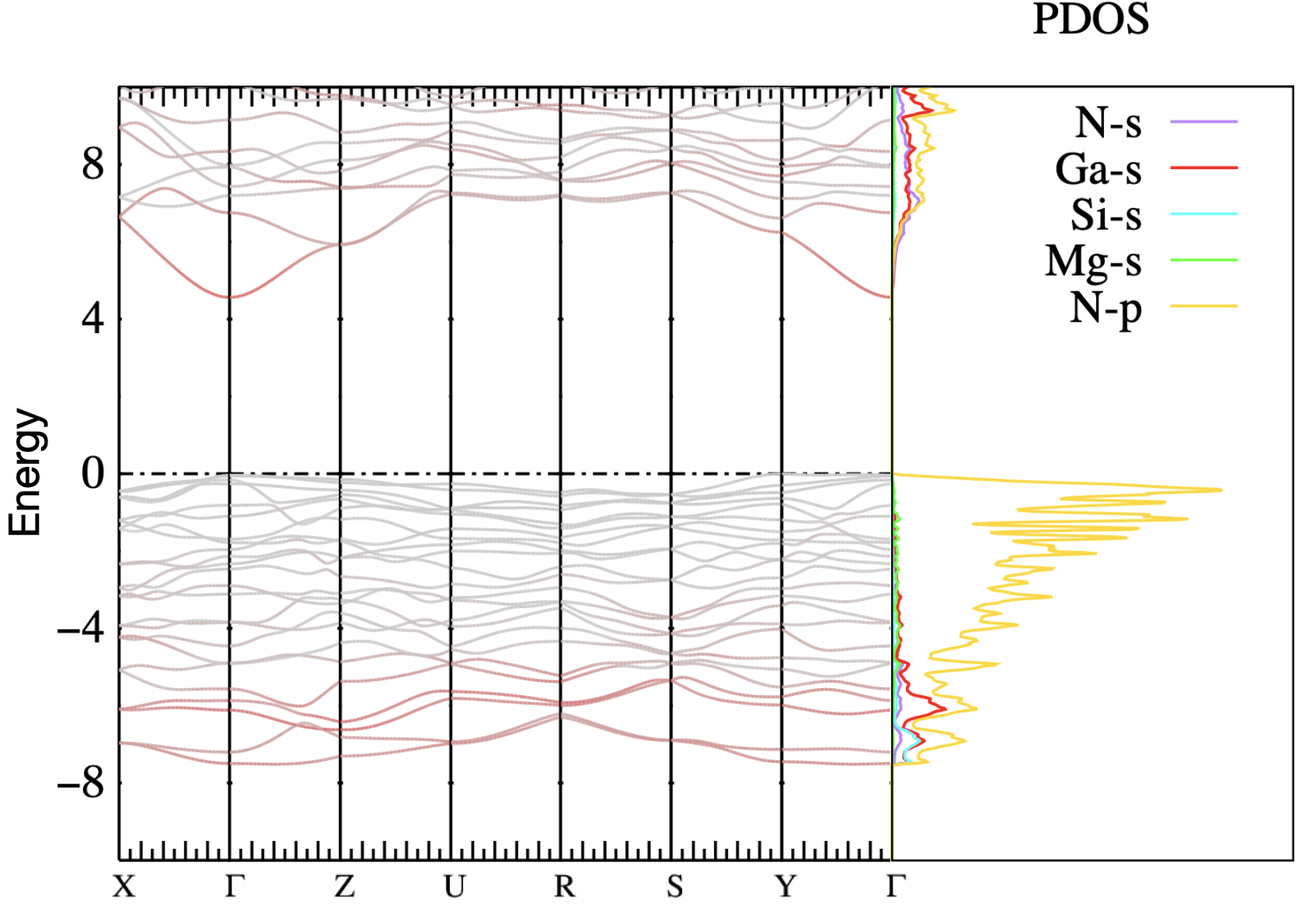}
      \caption{Band structures and corresponding partial density of states are shown for
               both $Pmn2_1$ (top) and $P1n1$ (bottom) space group of MgSiGa$_2$N$_4$
               alloy. The band structures are calculated in the $0.8\Sigma$ approximation,
               and Ga-$s$ orbitals are color coded with red-scale while the rest of the
               orbitals are in gray-scale. In both structures, the VBM is found to be on
               the k-point $Y$, but the difference between the VBM and highest valence
               band energy at $\Gamma$ is less than 0.1 eV (Table~\ref{table:gap}). 
               Partial density of states (in arbitrary but consistent units of states/cell
               per eV) are given for the orbitals of interest. The valence bands are
               dominated by the N-$p$ orbitals (color gold), while the CBM consists of
               mostly Ga-$s$ orbitals (color red).\label{fig:pmn21}}
    \end{figure}

  Subsequently, we calculated their electronic band structures both in GGA and in the
  quasiparticle self-consistent $GW$ approximations. The results of the band gaps are
  included in Table~\ref{table:gap}. The band gaps here do not include zero-point motion
  electron-phonon coupling corrections. In Ref. \onlinecite{Atchara16} these were
  estimated to be of order $-0.2$ eV. On the other hand, the overestimate of the volume by
  about 2 \% leads to an underestimate of the band gap. Using the band gap deformation potential
  $dE_g/d\ln V$ of -8.7 eV \cite{Atchara16},  the volume correction would increase the gap by about 0.2 eV.
  Thus these two corrections nearly cancel each other. For GaN, likewise, our gap here is underestimated by
  about 0.2-0.3 eV because of our use here of the GGA volume.  
  The band structures are shown in
  Fig~\ref{fig:pmn21}. The overview band structure shows that the top set of valence bands
  between $-8$ and 0 eV are predominantly N-$2p$ like as expected for a nitride. These
  show projections on the local  partial waves. Near the bottom of the valence band we can
  see that the states are bonding between N-$2p$ and Si-$s$ and Ga-$s$. The Mg-$s$
  contribution is less strongly present. The conduction band minimum atomic orbital
  character is more readily seen in the colored bands which indicate the contribution of
  the Ga-$s$ muffin-tin-orbital basis functions. The corresponding Si-$s$ and Mg-$s$ are
  not shown but are significantly smaller.  The predominant Ga-$s$ rather then Mg or Si-$s$ character of the
  CBM is consistent with a quantum well model. The $Pmn2_1$ is essentially an ultra-thin superlattice of
  half unit cell GaN  quantum wells and half unit cell MgSiN$_2$ units stacked along the {\bf b}-direction.
  The band offset is type I with a conduction band off set of about 1.4 eV according to Ref. \cite{Lyu2019} and hence
  if the quantum well just captures a single bound state, it could be at about 1.4 eV above the GaN CBM.
  The quantum confined effects for the holes are expected to be smaller since the effective masses are much larger.
  The wider the barrier of MgSiN$_2$ in this case, the stronger the quantum confinement effects would be and hence
  the larger the gap of the (MgSiN$_2$)$_x$(GaN)$_{1-x}$ alloys. 

    \begin{table}
      \caption{Band gaps of compounds and alloys in various approximations.
               \label{table:gap}}
      \begin{ruledtabular}
        \begin{tabular}{lcccc}
          Compound                   &          & GGA  & 0.8$\Sigma$   & 0.8$\Sigma$ + $\Delta(0)$\footnote{$\Delta(0)$ is an estimated zero-point motion correction from Ref. \cite{Atchara16}}\\\hline
                                                       
          MgSiN$_2$                  & Indirect & 4.03 & 6.15          & 5.84   \\
          MgSiN$_2$                  & Direct   & 4.36 & 6.52          & 6.28  \\
          GaN                        & Direct   & 1.72 & 3.08          &   \\
          MgSiGa$_2$N$_4$ ($Pmn2_1$) & Direct   & 2.98 & 4.83          &   \\
          MgSiGa$_2$N$_4$ ($Pmn2_1$) & Indirect & 2.92 & 4.76          &   \\
          MgSiGa$_2$N$_4$ ($P1n1$)   & Direct   & 2.73 & 4.60          &   \\
          MgSiGa$_2$N$_4$ ($P1n1$)   & Indirect & 2.70 & 4.57          &   \\
        \end{tabular}
      \end{ruledtabular}
    \end{table}

    \begin{figure*}
      \centering
      \includegraphics[width=8cm]{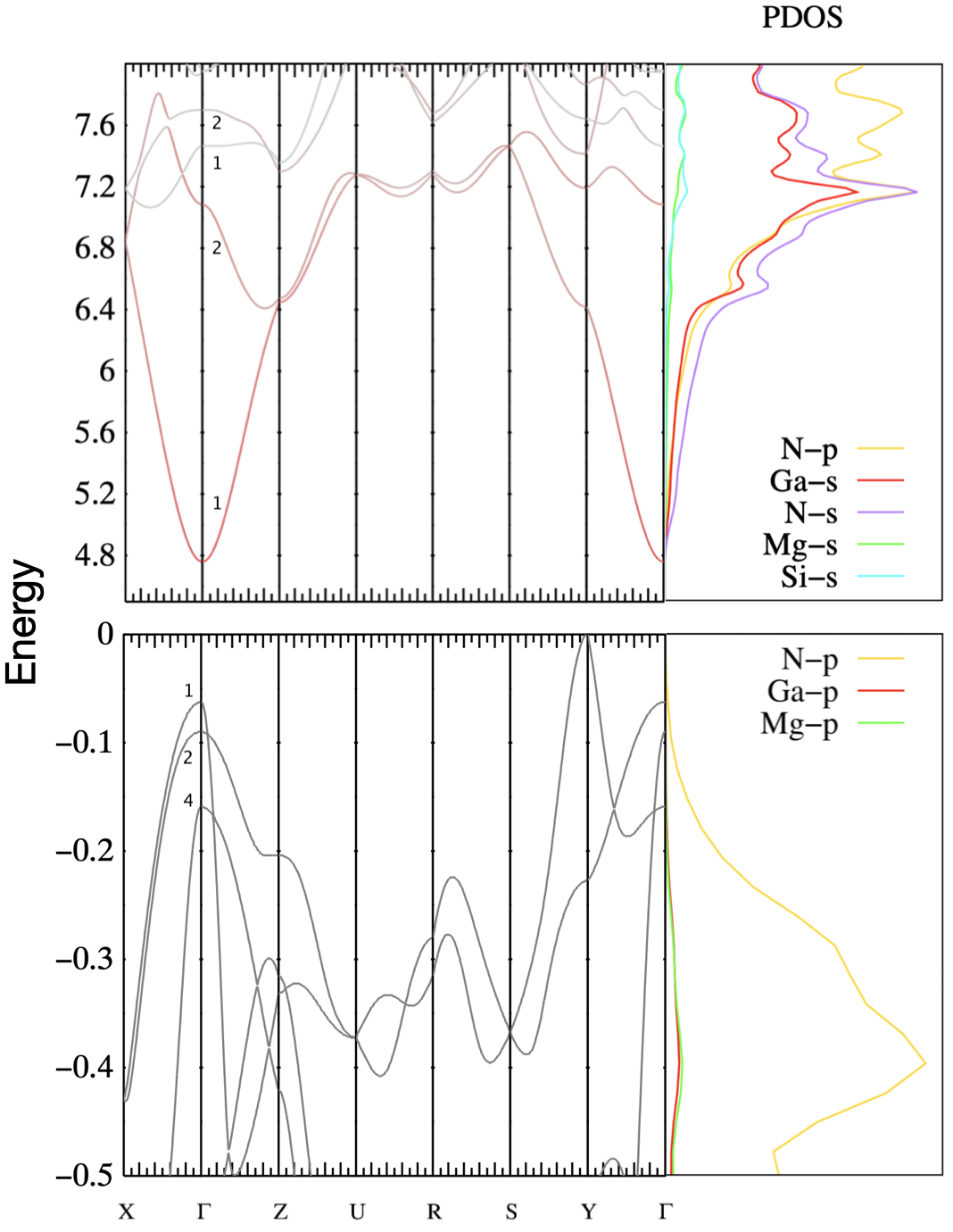}
      \includegraphics[width=7.55cm]{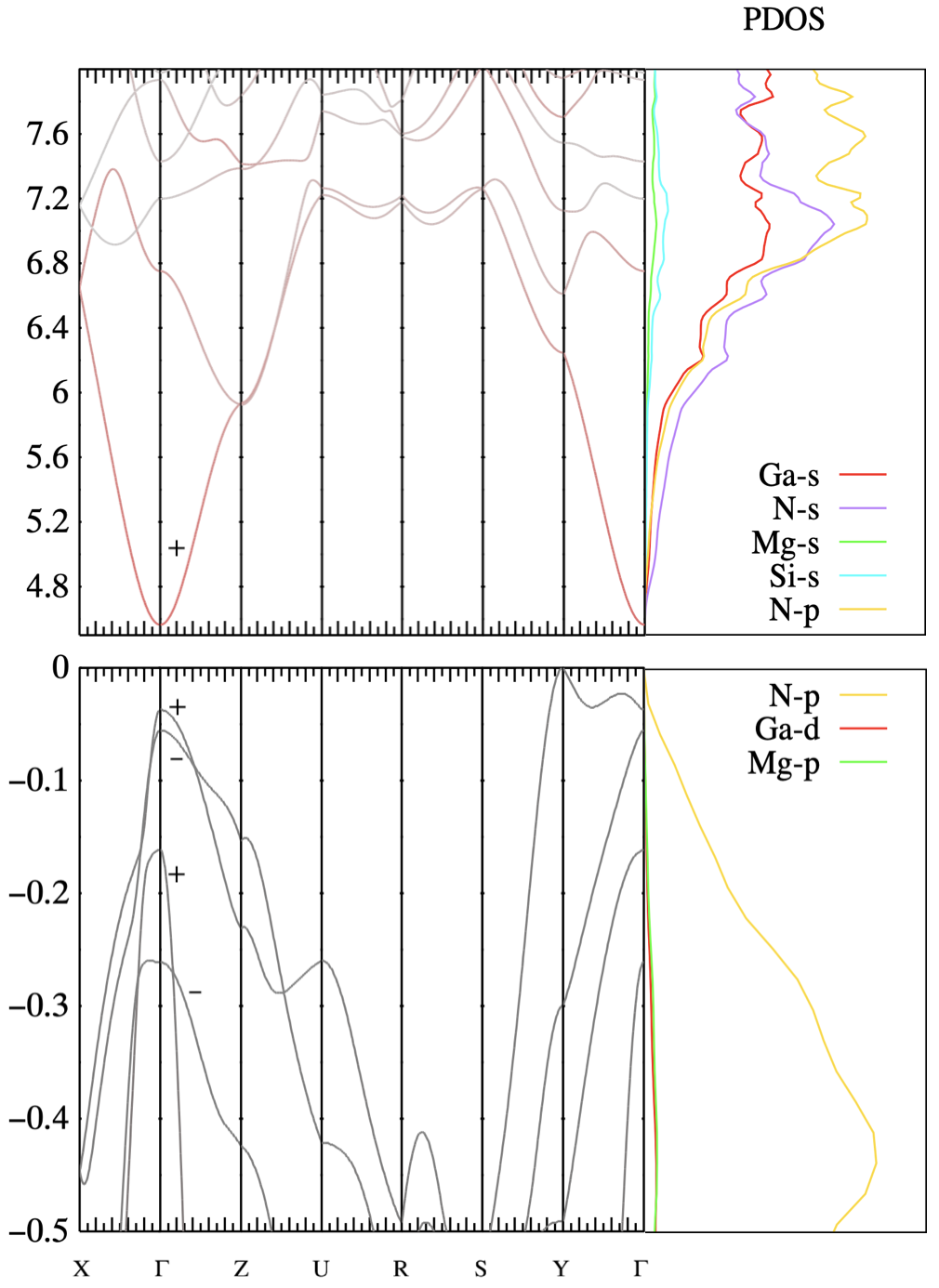}
      \caption{Band splitting and their symmetries are shown for $Pmn2_1$ (left) and
               $P1n1$ (right) structures, along with the partial density of states of
               dominant orbitals. Same color coding as in Fig.~\ref{fig:pmn21} is used.
               \label{fig:pmn21zoomc}}
    \end{figure*}

  A zoom in near the VBM and CBM for both structures shows that the VBM does not occur at
  $\Gamma$ but rather at $Y$, $(0,\pi/2b,0)$, and thus, strictly speaking, we still have an indirect gap
  system. However, this location of the VBM is different from that of MgSiN$_2$, where it
  occurs at $U$, $(\pi/2a,0,\pi/2c)$, and the energy difference between the VBM at $\Gamma$ and $Y$ here is very
  small. Furthermore, this is expected to be rather sensitive to strain. It also shows
  details of the crystal field splitting at $\Gamma$ in $Pmn2_1$ structure. The PDOS in
  the CBM shows that near the CBM the N-$s$ partial waves dominate  over the cation-$s$
  states but we should recall that these are anti-bonding $s$ states and that the dominant
  Ga-$s$ muffin-tin-orbital (as shown from the band color) extends into the N spheres and
  these thus also contribute to the N-$s$ PDOS.

  The bands at $\Gamma$  were symmetry labeled for $Pmn2_1$.  The point group is $C_{2v}$.
  In this point group, $\Gamma_1$ (or $a_1$) correspond to the $z$ basis function with
  $x,y,z$ along ($a$,$b$,$c$) respectively, $\Gamma_2$ corresponds to $b_2$ or $y$ and
  $\Gamma_4$ to $b_1$ or $x$. The $a_2$ or $\Gamma_3$ corresponds to $xy$ basis function.
  The CBM has $\Gamma_1$ symmetry and is $s$-like. Thus allowed dipole transitions occur
  between the VBM at $\Gamma$ and the CBM for ${\bf E}\parallel{\bf c}$. For
  ${\bf E}\parallel{\bf b}$ the transitions occur from the second VBM, and for
  ${\bf E}\parallel{\bf a}$ they occur for the third VBM. The valence band splittings are
  given in Table~\ref{table:split}.
  We may also note that the $z$-like $\Gamma_1$ state has the smallest mass along the
  $\Gamma-Z$ direction, while the next $\Gamma_2$ or $y$-like state has the smallest mass
  in the $\Gamma-Y$ direction and the $\Gamma_4$ or $x$-like state has its lowest mass in
  the $\Gamma-X$ direction. Unfortunately, the top of the VBM is thus $z$-like which
  implies that the same problem occurs as for Al$_x$Ga$_{1-x}$N at high Al-concentration
  $x$. Namely, the lowest energy transition is predominantly transverse magnetic (TM)
  which hinders light extraction for normal exit angle from the basal plane. However, this
  is sensitive to strain and the splitting between the $\Gamma_2$ and $\Gamma_1$ state is
  quite small and could perhaps more easily be reversed by some strain engineering.
  Also, it might be possible to grown these materials as films with other surfaces than the basal plane,
  in which case the {\bf c}-axis would be in the plane. 

  For the $P1n1$-structure, we have only a double-glide mirror plane $n$ perpendicular to
  the $b$-axis, so the point group is $C_s$ and states can only be labeled even or odd
  with respect to the mirror $n_y$ mirror-plane. Obviously, $y$ orbitals are odd with
  respect to this mirror-plane while $x$ and $z$ are even. We can see some interesting
  changes in the band structure reflecting the lower symmetry. First, at $\Gamma$, there
  are now four states instead of three within the first 0.3 eV below the VBM. Along the
  $\Gamma-X$ line we can see that the top two bands cross, indicating that they have
  different irreducible representations, respectively $+$ and $-$. The next band down has
  $+$ symmetry and therefore an avoided crossing with the $+$ band emanating from the VBM
  at $\Gamma$. Along $\Gamma-Y$ however, there is no symmetry left at all and hence the
  top bands can repel each other while in the $Pmn2_1$ these bands were allowed to cross
  because of their different even or odd character with respect to the $m_x$ mirror plane.

  We have also determined the effective mass tensors at the band edges, as given
  in Table \ref{table:meff}. The curvatures are determined by fitting the bands to a parabolic dispersion very close
  to the band edge in question in several directions and subsequently determining the principal values
  and direction of the constant energy surfaces. 
  The conduction band mass
  tensor shows only slight anisotropy  consistent with the point group symmetry. In the $P1n1$ case,  one
  principal axis is along $y$, the other two can be at some angle from $x$ in the $xz$ plane because the point group
  is only $C_s$.  For the CBM, we find these angles to be near 45$^\circ$. 
  The three highest valence bands at
  $\Gamma$ show a larger anisotropy with the smallest negative mass occurring in the direction that corresponds
  to the irreducible representation of the band in question.  In terms of the well known ${\bf k}\cdot{\bf p}$
  expression for the inverse effective mass tensor, 
  \begin{equation}
    M^{-1}_{\alpha\beta}= \delta_{\alpha\beta}\frac{1}{m_e} +\frac{1}{m_e^2}\sum_{n^\prime\ne n}
    \frac{\langle n{\bf k}|p_\alpha|n^\prime{\bf k}\rangle\langle n^\prime{\bf k}|p_\beta|n{\bf k}\rangle +c.c.}{E_{n{\bf k}}-E_{n^\prime{\bf k}}}
  \end{equation}
  we can see that the interaction of the top valence band with the conduction band would give a small negative contribution to
  the $M^{-1}_{zz}$ and none at all for $M^{-1}_{xx}$ or  $M^{-1}_{yy}$ because only $p_z$ matrix elements can couple the VBM of $\Gamma_1$ symmetry
  to the $\Gamma_1$ CBM.  On the other hand in the $x$ ($y$) directions, there are valence band states just below it
  of symmetry $\Gamma_4$ ($\Gamma_2$) which will couple to the VBM of $\Gamma_1$ symmetry and hence give a 
  large positive change in $M^{-1}_{xx}$ or $M^{-1}_{yy}$. This means a large negative change in the mass itself because
  $\Delta (1/m)=-\Delta m/m^2$. Thus we find strongly negative 
  masses for the $x$ and $y$ directions for the $\Gamma_1$ VBM. This implies a high positive hole mass in these directions.
  Other lower lying $\Gamma_1$ states will also give a negative
  contribution to the effective mass in the $z$ direction but, because these states are farther away, they will result in a smaller negative mass
  in the $z$ direction.   Similar reasoning explains the other cases. For the $\Gamma_2$ second valence band, the $\Gamma_1$ state
  above it which is the smallest energy difference, will give a strong negative contribution to $M^{-1}_{yy}$ in the $\Gamma-Y$ direction
  so a positive change to the mass itself which must compensate the effect from bands of $\Gamma_1$ symmetry lying deeper in the valence band.
  Hence the smallest mass along the $y$ direction for the state of $\Gamma_2$ or $y$ symmetry. 
  For the $P1n1$ case, the top valence band along $y$ is seen to have a positive curvature because of the strong repulsion
  of the two two valence bands along this symmetry line.
  For the mass tensor at $Y$ we can see for $Pmn2_1$ that $(m_xm_ym_z)^{1/3}=1.09$, so the effectively average density of states mass
  is close the free electron mass. For the $P1n1$ structure the effective density of states mass is 1.48, thus somewhat higher.
  While the conduction band masses are similar and again nearly isotropic. 
  
    \begin{table}
      \caption{Valence-band splittings (in meV) at the $\Gamma$ point relative to actual
               VBM at $Y$.\label{table:split}}
      \begin{ruledtabular}
        \begin{tabular}{lccc}
          Symmetry    & Pmn2$_1$ & Symmetry   & P1n1 \\\hline
                 
          $\Gamma_1$  & -63      & $+$ & -37  \\
          $\Gamma_2$  & -90      & $-$ & -56  \\
          $\Gamma_4$  & -159     & $+$ & -162 \\
                      &          & $-$ & -261 \\
        
        \end{tabular}
      \end{ruledtabular}
    \end{table}

    \begin{table}
      \caption{Electron effective masses (in units of free electron mass $m_e$) at
               $\Gamma$ for CBM and top valence bands in close vicinity to VBM are given
               for $Pmn2_1$  and $P1n1$ structures. The mass tensor components at the true VBM at $Y$ are  also
               given. The negative values for valence bands indicate positive hole
               masses. For the orthorhombic case ($Pmn2_1$) , the principal axes are along the
               crystal axes $x,y,z$ corresponding to $a,b,c$; for the $P1n1$ monoclinic space group, one principal axis is along $y$, the others in the $xz$ plane make an angle $\alpha_x$ from the $x$ axis as indicated.\label{table:meff}}
      \begin{ruledtabular}
        \begin{tabular}{llrrccc}
                & CBM   & $\Gamma_1$ & $\Gamma_4$ & $\Gamma_2$ & $Y$      \\\hline
                                                                          
          $m_x$ & 0.244 &  -3.059    & -0.224     & -4.321     &  -0.714  \\
          $m_y$ & 0.257 & -11.302    & -3.887     & -0.272     &  -1.105  \\        
          $m_z$ & 0.226 &  -0.196    & -2.567     & -2.494     &  -1.666  \\\hline\hline          
                & CBM   & $+_1$      & $-_1$      & $+_2$      & $-_2$   & $Y$     \\\hline
                                                                                   
          $m_1$     & 0.243      & -0.286     & -0.459     &  -0.196     & -2.372   & -0.664        \\
      $\alpha_x$    & -43.6       & 2.5       &  -3.7      &   91.2      & -87.8    &  0.5 \\
          $m_2$     & 0.240      &  -2.988    &   -3.936   &  -2.652     &  -2.652  & -1.627 \\
      $\alpha_x$    & 46.4       & 92.5       &  86.3      &   1.2       & 2.2      & 89.5\\
          $m_y$     & 0.264      & 2.167      &   -0.948   & -2.348      &  -0.251  &  -3.025 \\
        \end{tabular}
      \end{ruledtabular}
    \end{table}

\begin{figure}
      \includegraphics[width=\linewidth]{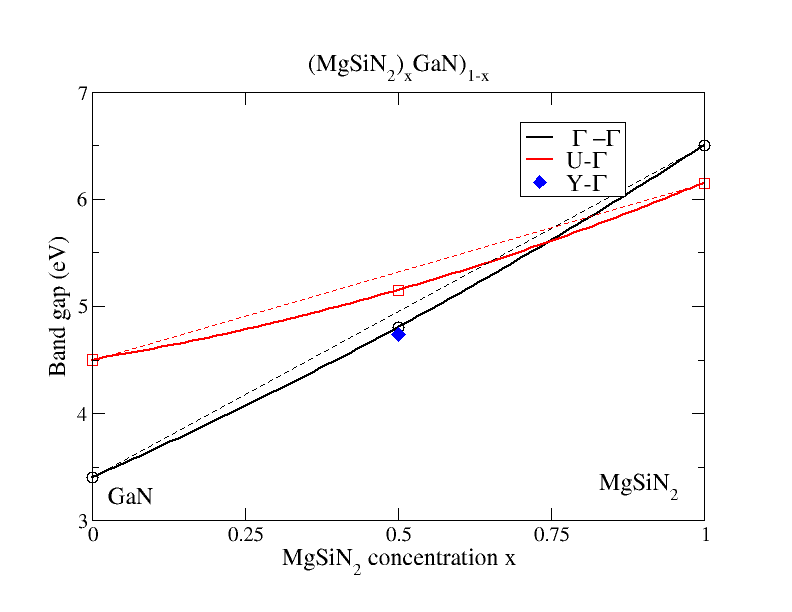}
      \caption{Band gaps in the (MgSiN$_2$)$_x$(GaN)$_{1-x}$ alloy system.\label{gapalloy}}
            \end{figure}

    Finally, we make an estimate of the band gap bowing in the alloy system. The band gap is estimated
    as $E_g(x)=E_g(0)+xE_g(1)-bx(1-x)$. For the direct $\Gamma-\Gamma$ gap we find $b\approx0.6$ eV, while
    for the $U-\Gamma$ gap is it about $0.7$ eV.  The results are shown in Fig. \ref{gapalloy}. They show
    that a band crossing between direct and indirect gap still occurs at about $x=0.75$ with a gap as high as 5.6 eV
    at this point.    We note that in principle,  octet-rule conserving  superlattice type structures with
    half unit cells of MgSiN$_2$ and GaN stacked along the $b$ direction as building blocks can be designed.\cite{Dinushi18} However, in a more disordered alloy system, one may expect some octet-rule violating motifs to exist, in particular
    those that involve a change in local valence by only $\pm1$. Their presence would likely decrease the gap
    by creating a tail of defect related states in the gap near the band edges. From related ternary compounds such
    as ZnGeN$_2$ we expect these to reduce the gap by about 0.5$\pm0.2$ eV. This would still leave a nearly direct gap larger than 4 eV but with a larger band gap bowing of $b\approx2.6$. In such a more disordered system, the distinction
    between the slightly lower indirect gap $Y-\Gamma$ and the direct gap would become essentially wiped out by
    the disorder induced band broadening.

\section{Conclusions}
  We have shown that a 50 \% alloy of MgSiN$_2$ and GaN  is promising as an alternative to
  Al$_x$Ga$_{1-x}$N alloy for increasing the gap beyond that of GaN to allow for
  opto-electronic applications into the deep UV region. First, we find that two
  octet-ruling compounds  exist with the composition MgSiGa$_2$N$_4$ which have negative
  energies of formation  and only a small positive mixing energy with respect to GaN and
  MgSiN$_2$. While we expect octet-rule violating motifs to increase the mixing energy
  we still expect such alloys to be feasible. 
  In its lowest energy structure with space group $Pmn2_1$,  the lattice
  constant in the $a$ direction has only a $+0.5$ \% mismatch with the corresponding
  lattice constant in GaN and the positive sign indicates that on a GaN substrate this
  compound would be under compression in this direction. In the other direction in the
  basal plane, its mismatch is $-2.3$\%. The band gap of this structure is nearly direct,
  by which we mean that the direct gap is less than 0.1 eV below the direct gap at
  $\Gamma$, and the direct gap is calculated to be 4.83 eV. Another structure with space
  group $P1n1$ also has fairly good lattice mismatch from GaN and a nearly direct gap of
  4.76 eV. These small deviations from direct gap character could possibly be overcome by
  strain tuning. 
   
  We should point out some caveats here. For a quaternary system, achieving the precise
  $Pmn2_1$ ordering is even more challenging than for a ternary and going away from the
  perfect 50\% stoichiometry of MgSiN$_2$-(GaN)$_2$ would likely introduce even more
  cation disorder in the form of locally octet-rule violating motifs.
  This is expected to decrease the band gap somewhat but a direct gap larger than
  4 eV is still expected.  On  the other hand, octet-rule satisfying superlattices at different
  concentrations are in principle possible and may provide higher gaps but are even more challenging to
  achieve. 
  To facilitate comparison with future experimental realization of these materials we have
  provided  ample detail on their predicted crystal structure and band structure, including  the effective mass
  tensors at the band edges.

\acknowledgements{The calculations were performed on the High Performance Computing
  Resource in the Core Facility of Advanced Research Computing at Case Western Reserve
  University. This work was supported by the US Department of Energy Basic Energy Sciences
  (DOE-BES) under grant number DE-SC0008933.}\\

\appendix

\section{Structural detail} \label{app-struc}
The reduced coordinates of the atoms and their Wyckoff designation are given in Tables \ref{table:wyck1},\ref{table:wyck2}
for $Pmn2_1$ and $P1n1$ respectively.
\begin{table}[h!]
      \caption{Reduced coordinates of atoms for the $Pmn2_1$ structure.
               \label{table:wyck1}}
      \begin{ruledtabular}
        \begin{tabular}{lcccc}
          Atom              & Wyckoff$^{a}$ & $x$     &  $y$      & $z$     \\
          \hline
          Ga                & 4b            & 0.25162 & -0.32950  & 0.49925 \\
          Mg                & 2a            & 0.00000 &  0.16007  & 0.49196 \\
          Si                & 2a            & 0.00000 & -0.17887  & 0.00043 \\
          N$_{\mathrm{Ga}}$ & 4b            & 0.22286 & -0.32754  & 0.88188 \\
          N$_{\mathrm{Mg}}$ & 2a            & 0.00000 &  0.13495  & 0.90272 \\
          N$_{\mathrm{Si}}$ & 2a            & 0.00000 & -0.19564  & 0.34264 \\         
        \end{tabular}                                    
      \end{ruledtabular}
      \begin{tablenotes}
        \small
          \item $^{a}$ $2a$ positions are ($x$,$y$,$z$) and (1/2,$-y$,$z+$1/2),
                $4b$ positions are ($x$,$y$,$z$), ($-x+1/2$,-y,$z+1/2$),
                ($x+1/2$,-y,$z+1/2$), and ($-x$,$y$,$z$).
      \end{tablenotes}      
    \end{table}

    \begin{table}[h!]
      \caption{Reduced coordinates of atoms for the $P1n1$ structure.\label{table:wyck2}}
      \begin{ruledtabular}
        \begin{tabular}{lcccc}
          Atom                & Wyckoff$^{a}$ &  $x$     &  $y$      &  $z$     \\
          \hline
          Ga$_{\mathrm{1}}$   & 2a            &  0.49984 & -0.15530  & -0.00064 \\
          Ga$_{\mathrm{2}}$   & 2a            &  0.74905 & -0.32606  &  0.49904 \\
          Mg                  & 2a            &  0.25124 & -0.33223  &  0.49367 \\
          Si                  & 2a            &  0.00024 & -0.16716  &  0.00067 \\
          N$_{\mathrm{Ga1}}$  & 2a            &  0.50965 & -0.13049  &  0.38153 \\
          N$_{\mathrm{Ga2}}$  & 2a            &  0.26846 &  0.30852  &  0.38142 \\
          N$_{\mathrm{Mg}}$   & 2a            &  0.73140 &  0.32936  &  0.40271 \\
          N$_{\mathrm{Si}}$   & 2a            & -0.00986 & -0.16526  &  0.34159 \\        
        \end{tabular}                                    
      \end{ruledtabular}
      \begin{tablenotes}
        \small
          \item $^{a}$ All atoms in $2a$ positions: ($x$,$y$,$z$), ($-x+1/2$,-y,$z+1/2$)
      \end{tablenotes}      
    \end{table}

\bibliography{Bib/dft,Bib/lmto,Bib/gw,Bib/zgn,Bib/zsn,Bib/mgn,Bib/msn,Bib/algan}
\end{document}